\documentclass[10pt, conference]{IEEEtran}
\IEEEoverridecommandlockouts
\usepackage{cite}
\usepackage{amsmath,amssymb,amsfonts}
\usepackage{algorithmic}
\usepackage{graphicx}
\usepackage{textcomp}
\usepackage{xcolor}
\usepackage{multirow}
\usepackage{array}

\def\BibTeX{{\rm B\kern-.05em{\sc i\kern-.025em b}\kern-.08em
    T\kern-.1667em\lower.7ex\hbox{E}\kern-.125emX}}
\begin{document}

\title{
A Comparative Study on Reward Models for UI Adaptation with Reinforcement Learning

% A Comparative Study on Reward Modelling for RL-based Adaptive User Interfaces

%A Comparative Study of Reinforcement Learning Techniques for Adaptive User Interfaces
%Using Reinforcement Learning Human Feedback for User Interface Adaptation: An Exploratory Study
}

%%
%
%   
%
%%
%An Exploratory Study on Reinforcement Learning Human Feedback for User Interface Adaptation

\author{
\IEEEauthorblockN{Daniel Gaspar-Figueiredo}
\IEEEauthorblockA{ITI \& Universitat Politècnica de València\\
Valencia, Spain \\
Email: dagasfi@epsa.upv.es}\\ 
\IEEEauthorblockN{Marta Fernández-Diego}
\IEEEauthorblockA{Universitat Politècnica de València\\
Valencia, Spain\\
Email: marferdi@omp.upv.es}
\and
\IEEEauthorblockN{Silvia Abrahão}
\IEEEauthorblockA{Universitat Politècnica de València\\
Valencia, Spain\\
Email: sabrahao@dsic.upv.es}\\ 
\IEEEauthorblockN{Emilio Insfran}
\IEEEauthorblockA{Universitat Politècnica de València\\
Valencia, Spain\\
Email: einsfran@dsic.upv.es}
}

\maketitle

\begin{abstract}

\textit{Background:}  
Adapting the User Interface (UI) of software systems to user requirements and the context of use is challenging. The 
main difficulty 
consists of suggesting the right adaptation at the right time in the right place in order to make it valuable for end-users. We believe that recent progress in Machine Learning techniques provides useful ways in which to support adaptation more effectively. In particular, Reinforcement learning (RL) can be used to personalise interfaces for each context of use in order to improve the user experience (UX). However, determining the reward of each adaptation alternative is a challenge in RL for UI adaptation. Recent research has explored the use of reward models to address this challenge, but there is currently no empirical evidence on this type of model.
\textit{Objective:} In this paper, we propose a confirmatory study design that aims to investigate the effectiveness of two different approaches for the generation of reward models in the context of UI adaptation using RL: (1) by employing a reward model derived exclusively from predictive Human-Computer Interaction (HCI) models (HCI), and (2) by employing predictive HCI models augmented by Human Feedback (HCI\&HF).
\textit{Method:} The controlled experiment will use an AB/BA crossover design with two treatments: HCI and HCI\&HF. We shall determine how the manipulation of these two treatments will affect the UX when interacting with adaptive user interfaces (AUI). The UX will be measured in terms of user engagement and user satisfaction, which will be operationalized by means of predictive HCI models and the Questionnaire for User Interaction Satisfaction (QUIS), respectively.
By comparing the performance of two reward models in terms of their ability to adapt to user preferences with the purpose of improving the UX (i.e. increasing user engagement, improving user satisfaction), our study contributes to the understanding of how reward modelling can facilitate UI adaptation using RL.

% [OLD ABSTRACT]

% \textit{Background.} Adaptive systems and user interfaces are becoming increasingly important in modern software applications. RL is a type of machine learning that can be used to personalise user interfaces for each context of use to improve user experience. One challenge of RL for UI adaptation is determining the reward of each adaptation. Recent research has explored the use of Reward Models (RM) to address this challenge. In this paper, we propose an exploratory study design to compare the effectiveness of using a reward model derived from predictive HCI models exclusively (HCI) and helped with human feedback (HCI\&HF). 
%
%\textit{Objective.} Our study aims to investigate the effectiveness of these different approaches for generating RMs in the context of UI adaptation using reinforcement learning. 
%By comparing the performance of different methods in terms of their ability to adapt to user preferences, improve user experience, and increase user engagement, our study contributes to the understanding of how reward modeling can facilitate UI adaptation using reinforcement learning.
%
% \textit{Method.} The experiment will use an AB/BA crossover design with two treatments: HCI and HCI\&HF. The study will employ the use of HCI models to obtain the user engagement and the SUS questionnaire to obtain the user satisfaction. We will analyse the data using statistical methods, including hypothesis testing and effect size calculation

% \textit{Results.} This is the Results
% \textit{Conclusion.} This is the Conclusions

\end{abstract}


    \begin{figure*}[]
        \centering
        \includegraphics[width=0.99\linewidth]{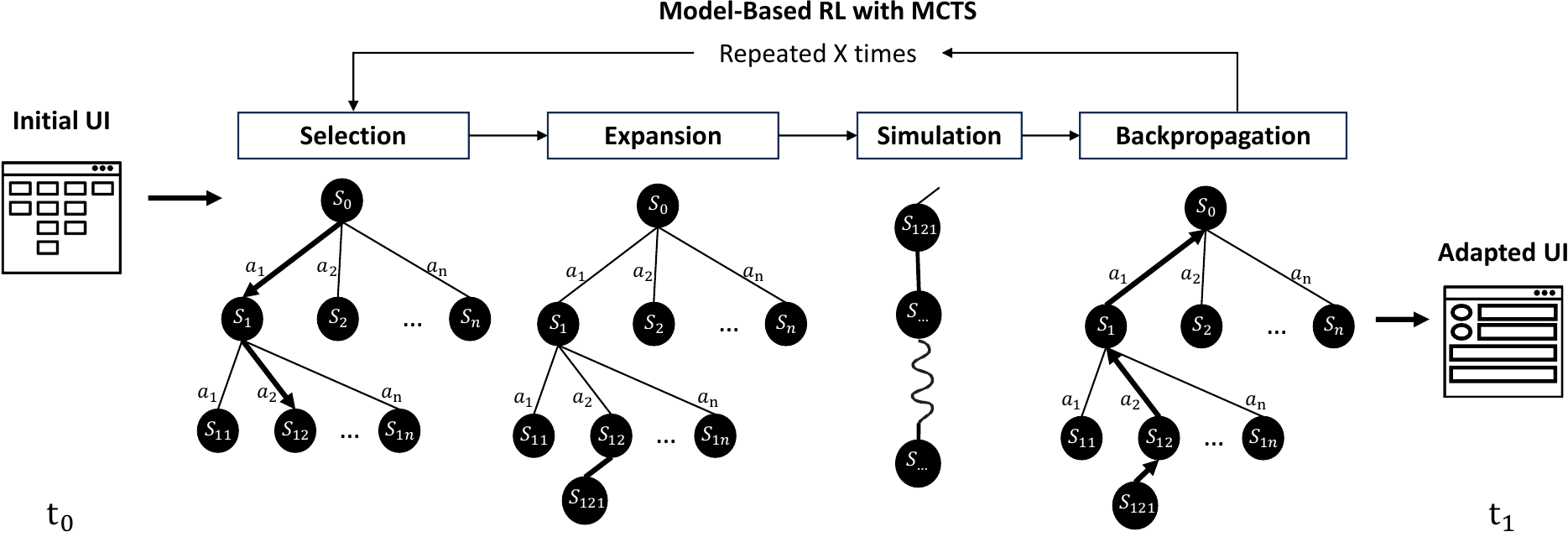}
        \caption{An interface is adapted by simulating several possible sequences of adaptations and evaluating them using predictive models in HCI. This approach avoids greedy, disadvantageous adaptations, and may anticipate possible user responses even with limited observation data. Figure adapted from~\cite{Chaslot:2021}}
        \label{fig:MCTSapproach}
    \end{figure*}

\section{Introduction}

Adaptive systems and adaptive user interfaces (AUI) have become increasingly important in modern software applications. They have been introduced in order to address some of the usability problems in many software applications~\cite{Akiki:2014}. These systems can change aspects of their structure, functionality, or interface to accommodate the differing needs of individuals or groups of users and the changing needs of users over time~\cite{Viano:2000}. 
%Self-adaptive user interfaces have been promoted as a solution to deal with context variability due to their ability to automatically adapt the UI to the context-of-use (i.e., platform, environment or user) at run-time. 
However, suggesting the right adaptation at the right time in the right place in order to make it valuable for end-users is challenging.
%~\cite{Yigitbas:2020}. 
%The development of this type of approaches is a challenging and complex task as additional aspects like context management and UI adaptation have to be covered~\cite{Yigitbas:2020}.
%Recent progress in ML techniques provides useful ways in which to support UI adaptation more effectively. 
%Research on adaptive user interfaces is moving towards the use of machine learning~\cite{Nesrine:2013}. %% añadir mas refs, que las hay. 
In previous work, we proposed a framework for intelligent UI adaptation~\cite{abrahaoModel:2021}. This framework proposes the use of Machine Learning algorithms in order to provide valuable UI adaptations. 
% Falta poner una frase para decir porque hemos selecionado RL en el contexto de este trabajo. 
Several approaches have been proven successful in simple adaptation problems, such as recommendations and the calibration of interface parameters~\cite{Lomas:2016, Dudley:2019}.
However, Reinforcement Learning (RL) is more appropriate as regards learning policies for sequences of adaptations in which rewards are not immediately achievable.
Indeed, the UI adaptation problem can be defined as a stochastic sequential decision problem, where the adaptive system should plan a sequence of adaptations over a long horizon~\cite{Todi:2021MCTS}.
% [Reinforcement learning (RL) is a type of machine learning method appropriate for this type of problems.] 
RL involves training an agent to make decisions based on rewards received from the environment for certain actions~\cite{sutton:2018, kaelbling:1996}. In UI adaptation, the agent will make decisions regarding which adaptations to apply by considering the user's interaction behaviour. The goal of UI adaptation is to personalize the UI for each context of use so as to improve the user experience (UX). 

In this context, one challenge of RL for UI adaptation is determining the reward of each adaptation. Traditional RL uses numerical rewards, which are difficult to define in scenarios in which the outcome of the action or the agent’s goal is complex. 
%, but in complex scenarios where the outcome of an action is unpredictable it is very difficult to define a reward function with numerical values. 

Recent research has, therefore, explored ordinal rewards as an alternative~\cite{Zap:2020}, which induce scale-invariance and reduce manual reward engineering~\cite{Schmidt:2019}.
% Recent research has, therefore, explored the use of ordinal rewards as an alternative~\cite{Zap:2020}. 
% % This approach
% Ordinal rewards 
% have been shown to induce scale-invariance and reduce the need for manual reward engineering~\cite{Schmidt:2019}. 
Another 
%complementary 
approach is to make the reward function part of the learning process. In complex environments, the use of reward models can be very profitable. Reward models are a key component of RL that are used to specify the user's goals and the system~\cite{Chen:2022}. During the process of learning the goal, putting the human into the loop may be particularly powerful. Human Feedback (HF) could be used as regards specifying the goal more intuitively and quickly when compared to manual objective hand-crafted methods~\cite{Armstrong:2021, openaiDeepmind:2017}.

On the one hand, reward models in UI adaptation can be defined using predictive HCI models. 
An HCI model is a computational model that can explain how users interact with interfaces at the level of individual human cognition~\cite{Paton:2021}. It simulates consequences – benefits and costs – of possible adaptation sequences without actually executing them.
%ESTAS REFERENCIAS SON DE REINFORCEMENT LEARNING!!!!
%~\cite{sutton:2018, kaelbling:1996}.
%predicts how humans interact with computer interfaces. 
These models are based on various factors, such as the user's behaviour, task difficulty, and interface design features.
By simulating human behaviour, HCI models can provide insights into how users will interact with a new interface (or an adapted version) before it is even built, thus allowing software developers to make informed decisions about interface design. 
However, finding the best adaptation is computationally costly, especially when considering sequences of changes over a long horizon. To solve this computational problem in an online setting, Monte Carlo Tree Search (MCTS) could be used for planning adaptations~\cite{Todi:2021MCTS} (see Section 2). 
On the other hand, putting the user into the loop, or using user feedback, involves collecting user ratings or preferences on various aspects of the UI, which can be used to enhance a reward model by reflecting the users' goals and needs (preferences).
This falls in the promising field of Reinforcement Learning from Human Feedback (RLHF), or RL from human preferences, in which the HF helps to train a reward model~\cite{stiennon:2020, openaiDeepmind:2017}.
Although these techniques for generating reward models appear to be useful, there is, to the best of our knowledge, no empirical evidence concerning which is most effective as regards supporting UI adaptation in a given context of use.
%there is no empirical study which compares them to determine which one best supports UI adaptation in a given context of use.

In this paper, we present a confirmatory study design to compare two different approaches to generate reward models in the context of UI adaptation using RL.
Our study aims to investigate the effectiveness of using a reward model derived exclusively from predictive HCI models and predictive HCI models augmented with HF. 
The user goal may differ depending on the application domain. For example, user engagement is a key factor for the success of e-commerce systems and online catalogues as it can 
significantly 
impact on the users' purchase decisions and improve overall satisfaction. 
% In an online store, planning adequate adaptations of the UI may increase the customer engagement.
%
By comparing the performance of different types of reward modelling in terms of their ability to adapt to user preferences and improve UX, our study contributes to the understanding of how reward modelling can facilitate UI adaptation using RL.

\section{Background and Related Work}

Adaptive User Interfaces (AUI) have attracted increasing attention in recent years as a means to improve UX and task performance. Several studies have explored the benefits of using AUI in various application domains, such as e-commerce, education, and healthcare systems. The major challenge in this context is to determine the sequence of adaptations that have to be carried out in order to improve the system's quality, the user performance or UX.
To this end, several studies have applied different methods, such as defining rules and heuristics or the use of Machine Learning and RL methods. For example, 
in the context of menu searching in UI, MCTS has been proposed as a promising technique for the development of adaptive menu search interfaces \cite{Todi:2021MCTS}. 
%
% [Falta introducción a TODI: que es audes medaptación de IU para tarea de busqueda de menus y que utilica MCTS.]
% 
In this recent work, Todi et al.~\cite{Todi:2021MCTS} used predictive HCI models to predict rewards for each state during simulations. 
%simulate rewards for each state during rollouts. 
Since online simulations can be computationally expensive, a pretrained value network was used to directly obtain value estimates for unexplored states.
Training data for this neural network was generated using the predictive HCI models. The authors showed that while the computation time increases drastically with simulations as search depth increases,it remains constant with the neural network approach without interfering much in the overall success rate (92.7\% with model-based simulation \textit{vs.} 89.6\%).

% Figure \ref{fig:MCTSapproach} shows how MCTS planning could be applied to our domain. We start from an initial user interface, with some specific properties such as layout configuration, font sizes, colour schemes, 
% and the RL agent, through MCTS using HCI models for simulations, is able to determine 
% %and through MCTS, which use HCI models, the algorithm is able to determine 
% the sequence of actions that would lead to an improved user experience. 
%
%
%
%%%%%%%%%%%%%%%%%%%%%%%%%%%%%%%
Fig. \ref{fig:MCTSapproach} shows how MCTS planning could be applied to our domain. In the current state of the tree at \textit{t0} (i.e., the actual design of the interface and user observations from his/her previous interaction session with the interface), the algorithm \textit{i) selects} a child node based on a selection policy that balances exploration and exploitation until reaching the most promising leaf node; \textit{ii) expands} the node only if it has already been explored by adding child nodes for all the possible adaptations and pick one child node at random;
% by adding one or more child nodes to the tree if it has not been visited before;
\textit{iii) roll-outs} and simulates rewards using predictive HCI models while adaptations are chosen at random since the tree has no value estimates with which to inform the selection of consequential states. A sample of these roll-outs will be augmented by HF, and 
%, as the tree has no value estimates to inform the selection of consequential states, adaptations are chosen at random, and rewards are estimated using predictive HCI models;
\textit{iv) backpropagates} the updated statistics of all nodes visited during selection and expansion up the tree.
% the simulation up the tree to update the statistics of all nodes visited during selection and expansion.
These four steps are repeated until a stopping criterion is met.
% The final action chosen by MCTS is the one with the highest expected reward based on the statistics accumulated during search.
After the \textit{X} iterations allowed by the MCTS algorithm and a shallow roll-out horizon of \textit{H} steps, an adaptation decision is made on the basis of the highest expected reward from among the nodes that are located at \textit{t1}, resulting in an adapted UI. To move from \textit{t1} to \textit{t2}, the tree is re-unfolded from the current UI at \textit{t1}, and so on.
%%%%%%%%%%%%%%%%%%%%%%%%%%%%%%%%%%%%%%

Note that we have defined an environment in which to train and compare RL agents for the context of AUI in a previous study~\cite{gaspar-figueiredo:2023}. In this work, we have defined the \textit{state} representation, and the \textit{actions} that the system can perform to adapt the UI: i) change the UI layout; ii) change the colour scheme; iii) change the font size; iv) show/hide content, and v) do nothing.

In \cite{langerak:2022marlui}, the authors formulated UI adaptation as a multi-agent RL problem, with both a user agent who learns to interact with a UI so as to complete a task and an interface agent who learns UI adaptations to maximise the user agent’s performance. The joint exploration of both agents makes the adaptative UIs goal-agnostic.
However, we believe that HF could be used to optimise the reward model. For example, Christiano et al.~\cite{openaiDeepmind:2017}, propose a method with which to train RL agents using human preferences. The authors present an off-policy RL algorithm that learns from HF in the form of pairwise comparisons of trajectories. The algorithm uses a deep neural network to represent the value function and a ranking loss to optimise the network parameters. The authors evaluate their method with a variety of tasks, including Atari games and robotic manipulation, and show that it outperforms existing methods that use hand-crafted reward functions. The paper concludes that their method is a promising approach with which to train RL agents in complex environments where designing reward functions is difficult. In~\cite{schrittwieser:2021}, the authors propose an algorithm that does not require any special adaptations for the off-policy or offline RL settings.

% Fig. \ref{fig:ExecutionPlan} a) shows the entire process of training an RL agent which uses MCTS, with reward models such as HCI models and with human feedback. In our context, the RL agent would take actions (UI adaptations) and the environment will return observations. Then, a reward model will return reward predictions, based on 1) predictive HCI models only and 2) predictive models enriched with human feedback.

% y (Referencia al pa Then, per de TODI[model-based RL adaptainformation distribution.tion]) y se mostrara la dificultad de entrenar estos agentes por la falta de feedback o por tener un feedback incierto (Motivación para RLHF). Además de eso, se mostrarán los \textbf{algoritmos usados de RL} para adaptacion de IUs (principalmente \textbf{MCTS})

% \subsection{Reinforcement Learning}

% The human feedback has been used to reward reinforcement learning agents in diverse contexts [Citar algunos papers donde se use RLHF para chatbots, robotica, etc].

% Aqui podemos hablar de que es RLHF con más detalle y en qué ámbitos se ha aplicado

% Luego, mostraremos sus ventajas del RLHF: Principalmente Fine-tuning a lo que los usuarios Y sus inconvenientes: Feedback explicito.

% Deberemos mostrar tambien\textbf{ qué algoritmos de RL }usaron para entrenar los modelos, y si han usado datos iniciales de entrenamiento y cómo.

\section{Experimental Design}

According to the GQM template for goal definition~\cite{Basili:1994:GQM}, the goal of this study is to analyze \textit{reward models derived from predictive HCI models and predictive HCI models augmented with HF in RL algorithms} with the purpose to \textit{assess their impact to adapt UI} with respect to {their ability to improve the UX of software applications} from the point-of-view of  \textit{both developers and researchers interested in reward modelling in RL-based methods for UI adaptation}. The former might be interested in how the use of RL for UI adaptation can improve UX in order to inform the design of more effective interfaces. 
This study will also provide insights into obtaining user feedback for the development of adaptive systems that can effectively adapt the UI. The latter might be interested in our study results to delineate future research (e.g., means of attaining feedback with which to evaluate the UI adaptation). Furthermore, the findings could inform the development of new approaches and tools for the improvement of UX in software applications, and could ultimately lead to more efficient and effective software design and development processes. The context consists of a group of undergraduate and Master's degree students in Computer Science at the Universitat Politècnica de València interacting with UIs from the e-commerce and e-learning domains.

\subsection{Research questions and hypotheses}

\begin{figure}[]
        \centering
        \includegraphics[width=0.99\linewidth]{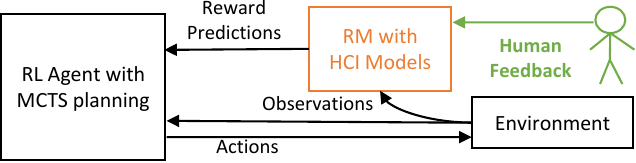}
        \caption{The RL agent employing two strategies to obtain reward predictions: predictive HCI models only (orange) and predictive HCI models with human feedback (green)}
        \label{fig:ExecutionPlan}
    \end{figure}

%In this section, we present the research questions and hypotheses that guide our investigation into the effectiveness of reward models for adapting UIs using RL. 
%We seek to understand how the use of reward models can facilitate UI adaptation using reinforcement learning and improve user experience. 
%In particular, 
We aim to study the systematic variation of reward models so as to measure their influence on the selection of UI adaptations that maximise the user's experience (i.e., user engagement and user satisfaction) in an experimental set-up. For this purpose, we shall implement and monitor two RL-based AUI strategies (using each of the reward models selected) in two different domains.
Fig. \ref{fig:ExecutionPlan} shows the context used to train the RL agent that employs MCTS, with reward models such as HCI models and with HF.
In our case, the RL agent will take actions (UI adaptations) and the environment will return observations. A reward model will then return reward predictions on the basis of 1) predictive HCI models only and 2) predictive models enriched with HF.

The objective of this study is to answer the following research questions and hypotheses:

% \begin{enumerate}

 % \item[$\textbf{RQ1}$] 
 $\textbf{RQ1}$: \textbf{Does the incorporation of HF in the reward models improve the effectiveness of AUIs in terms of user engagement, when compared to reward models derived exclusively from predictive HCI models?} RQ1 sheds light on the cost and benefits of training RL models with or without human feedback in terms of their impact on user engagement (operationalized by means of predictive HCI models). Since deciding about the best adaptation or sequence of adaptations to perform is not obvious for a human, incorporating humans into the loop to train the agent needs to be tested. As far as we are concerned, there is no evidence of the improvement in AUI.
According to literature, the use of human feedback could bring benefits such as capturing the way in which humans naturally propose or expect what to adapt, and when to make changes, but it could also have costs related to the ability of the system to scale with the availability of computational resources. Since the analysis of costs vs. benefits for the specific problem of AUIs obtained using RL models with or without human feedback has not yet been studied, we prefer to be conservative and not postulate an effect in favour of AUIs with human feedback. We therefore, propose the following null hypothesis:
    
         \begin{enumerate}
            \item[$H_{n1}$] There is no significant difference in user engagement between AUIs that incorporate HF in the reward models and those that use reward models derived exclusively from predictive HCI models.
        \end{enumerate}

% \end{enumerate}

Moreover, to assess the benefits of AUIs, it is desirable to understand the degree to which they improve the UX when compared to non-adaptive interfaces. This could be done by comparing the user engagement (operationalized using the User Engagement
Scale - UES~\cite{Obrien:2018}) and user satisfaction (operationalized using the Questionnaire for User Interaction Satisfaction - QUIS~\cite{norman:1998}) with the adapted and non-adapted UIs in different application domains.

 Literature has already demonstrated the positive effect of AUIs in certain contexts, but has also reported the limitations that could impede the expected benefits~\cite{Lavie:2010}: risk of misfit (the end-user’s needs are incorrectly captured or interpreted), user cognitive disruption (the end-user is disrupted by the adaptation), lack of prediction (the end user does not know when and how the adaptation will take place), lack of explanation (the end-user is not informed of the reasons for adaptation). Moreover, adopting a poor adaptation strategy may have negative effects on the user due to surprise or relearning effort. Such costly changes should be avoided. Thus, to be conservative, we prefer not to postulate that AUIs will be superior when compared to its non-AUIs version. Also, the impact of system adaptivity on user engagement has been studied in certain contexts (e.g., adaptive video streaming systems~\cite{Qiao:2021}, large display-based UIs~\cite{Li:2016}), but evidence to understand the benefits of using RL models to support UI adaptation is required. To achieve this, we formulate the following research questions and hypotheses:

%%%%%%%%%%%%%%%%%%%%%%%%%%%%%%%%%%%

$\textbf{RQ2}$: \textbf{Does the use of AUIs, obtained with reward models, improve \textit{user engagement} when compared to non-adaptive interfaces in different application domains?}

\begin{itemize}
    \item $H_{n2}$ There is no significant difference in user engagement between AUIs obtained with reward models derived exclusively from predictive HCI models and non-adaptive interfaces.

    \item $H_{n3}$ There is no significant difference in user engagement between AUIs obtained with reward models derived from predictive HCI models that incorporate HF and non-adaptive interfaces.
\end{itemize}

$\textbf{RQ3}$: \textbf{Does the use of AUIs, obtained with reward models, improve \textit{user satisfaction} when compared to non-adaptive interfaces in different application domains?}

         \begin{enumerate}
            %\item[$H_{n4}$] There is no significant difference in user satisfaction between adaptive and non-adaptive user interfaces.

            \item[$H_{n4}$] There is no significant difference in user satisfaction between AUIs obtained with reward models derived exclusively from predictive HCI models and non-adaptive interfaces.
            
            \item[$H_{n5}$] There is no significant difference in user satisfaction between AUIs obtained with reward models derived from predictive HCI models that incorporate HF and non-adaptive interfaces.

            % \item[$H_{n5}$] There is no significant difference in user satisfaction between adaptive user interfaces that incorporate human feedback in the reward models and those that use reward models derived exclusively from predictive HCI models.
        \end{enumerate}

    The scope of RQ1 lies in training the RL agent with the two different reward models, while the scope of RQ2 and RQ3 is the actual usage of systems (using UIs that do or do not employ the RL agent) by end users. The goal of the statistical analysis will be to reject these hypotheses and possibly to accept the alternative ones (\textit{e.g., }$H_{a1}$ = $\neg H_{n1}$). All the hypotheses are two-sided because we did not postulate that any effect would occur as a result of different reward models usage. 

\subsection{Variables}
% The independent variables are the factors that the researchers manipulate to test their effects on the dependent variables, which are the outcomes of interest. In this section, 

This section outlines the variables and 
%their abbreviations, 
measurement scales, and how they will be operationalized to ensure reliable data collection. Statistical analyses are detailed in Section III.E.
%the Analysis Plan section.

%In this section we present the independent and dependent variables that will be used in this study, as well as their operationalization and measurement. We provide a description of each variable, its abbreviation (if applicable), and its measurement scale. We also outline how each variable will be operationalized to ensure consistent and reliable data collection throughout the study. The specific dependent measures and statistical analyses is provided in the Analysis Plan section.

\subsubsection{Independent variables}

The main independent variable in this study is the type of reward model employed by the two RL-based UI adaptation strategies that will be implemented as part of our experimentation. The scale of this variable is nominal, and can assume two possible values: adaptive UI using predictive HCI models only (HCI) and adaptive UI using predictive HCI models augmented with HF (HCI\&HF).
However, to understand the gain in the improvement of UX when compared to a baseline (non-adaptive UI), we introduce a third system without any adaptive features (non-adaptive, NA). 
%The independent variable in this study is the type of User Interface Adaptation (UIA). There are three levels in
% of
%UI adaptation: non-adaptive UI (AUI-baseline), adaptive UI using predictive HCI models only (HCI), and adaptive UI using predictive HCI models and human feedback (HCI\&HF). Therefore, we use a nominal scale for this variable.
% 
% The predictive HCI models predict user engagement and inform the adaptive UI, whereas the human feedback provides additional input to HCI model prediction. 
The secondary independent variable is the application domain (i.e., adaptive system), which is a nominal variable with three possible values: Sports, Courses and Trips. The former is an online sporting goods store (e-commerce domain) while the latter is a Course management system (e-learning domain).

%Trip planning system (removed). 
%The UI adaptations are implemented in three different catalogues: A trips catalogue, a sports store catalogue, and a course catalogue. We will use the trips catalogue domain for the AUI-baseline. The sports and courses catalogue will be used for both HCI and HCI\&HF. The order of the catalogues is counterbalanced across participants.

%%%%%%
%%% No usaremos los "catalogos" como variable independiente.
% En caso de querer justificarlo, añadir esto o algo similar: 
% The use of different catalogues in this study is intended to reduce threats to [internal (?)] validity by testing the effectiveness of the adaptive UI methods without comparing their performance across different contexts. The focus of this study is not on comparing the effectiveness of the different catalogues, but rather on investigating the effectiveness of the reward models derived from predictive HCI models exclusively and with human feedback in improving user engagement in the context of adaptive UIs.
%%%%%%

\subsubsection{Dependent variables}

The dependent variable in this study is user experience (UX). It is defined by the ISO 9241-210: 2010 as \textit{"a person’s perceptions and responses that result from the use and/or anticipated use of a product, system or service}". In contrast to task-oriented interactions, UX represents a focus on “experience”, encompassing also  hedonic qualities, emotions and affect (e.g., interested, enthusiastic, irritable) that people experience while interacting with software systems. 
%We understand the relationship between UX and usability as the latter is subsumed by the former. 
%[x]. 
It can be operationalized in several ways as this is a complex concept with several dimensions. In this study, we measure UX in terms of user engagement and user satisfaction.

%The dependent variables in this study are \textsc{User Engagement} (UE), and \textsc{User Satisfaction} (US). 

%
%%%%%OLD%%%% 
%User engagement is the quality of the UX that emphasises the positive aspects of the interaction, and in particular the phenomena associated with being captivated by a software system, and so being motivated to use it~\cite{LEHMANN:2012}.
%
%It will be measured during the user interaction with the adaptive system (i.e., during the experimental task) using predictive HCI models that employs various metrics such as the number of clicks and time spent on each page, the frequency and number of user activities, recency, and the total number of actions performed during the user interaction. Using the study of ~\cite{LEHMANN:2012, barbaro:2020} as basis, these metrics will be used to calculate a variable in the ratio scale that ranges from 0 to 100\% of user engagement. Therefore, the definition of user engagement we use is based on previous studies from the HCI field ~\cite{LEHMANN:2012, barbaro:2020}.
%%%% NEW

%User engagement is the quality of the user experience that emphasises the positive aspects of the interaction, and in particular the phenomena associated with being captivated by a software system, and so being motivated to use it

User engagement is a quality of user experience with software systems that is characterised among others by aesthetic, sensory appeal, perceived control and time, awareness, motivation, and affect~\cite{obrien:2008}~\cite{doherty:2018}~\cite{LEHMANN:2012}. 
It will be measured in two different stages: 1) when training the RL model (for calculating reward predictions using predictive HCI models that employs behaviour logging approaches), and 2) during the user interaction with the system (i.e., experimental task). In the fist stage, user engagement will be operationalized using a set of metrics collected from the literature~\cite{LEHMANN:2012}, \cite{barbaro:2020}, \cite{carlton:2021} (i.e., individual event counts, relative frequencies of events, total number of user actions, and task completion time); while, in the second stage, it will be operationalized using the UES questionnaire in a post-experimental task. 
%These metrics have been used to measure user engagement in prior works ~\cite{LEHMANN:2012}, \cite{barbaro:2020}, \cite{ carlton:2021}. 

%MIRAR RSTO: It will be measured using predictive HCI models that employs behaviour logging approaches during 1) the training of the RL agent and 2) the user interaction with the adaptive system (i.e., experimental task). To this end, individual event counts, relative frequencies of events, the total number of user actions, and the time to complete each task will be obtained.

%However, in order to confirm that these measures are good predictors of user engagement, we will also measure this variable after the experimental task using the User Engagement Scale (UES) [x]. 
%The participants will be asked to fill in this questionnaire after the experimental task. 

%various metrics such as the number of clicks and time spent on each page, the frequency and number of user activities, recency, and the total number of actions performed during the user interaction. Using the study of ~\cite{LEHMANN:2012, barbaro:2020} as basis, these metrics will be used to calculate a variable in the ratio scale that ranges from 0 to 100\% of user engagement. Therefore, the definition of user engagement we use is based on previous studies from the HCI field ~\cite{LEHMANN:2012, barbaro:2020}.

%

% satisfaction %SEGUN LA ISO
%extent to which the user's physical, cognitive and emotional responses that result from the use of a system, product or service meet the user’s needs and expectations

On the other hand, user satisfaction is a measure of the quality of the UX and refers to the extent to which a user's expectations are met~\cite{iso9241:2018}. It will be measured after the user interaction with each of the adaptive systems (i.e., post-experimental task) using the Questionnaire for User Interaction Satisfaction (QUIS)~\cite{norman:1998}. It contains a measure of overall system satisfaction along six sub-scales and measures of four specific UI factors (screen factors, terminology and system feedback, learning factors, and system capabilities). Each area measures the overall satisfaction on a 10-point scale.

\subsection{Subjects}

 % The participants were chosen by means of convenience sampling. They attended the Fall 2017 course on Empirical Software Engineering with a focus on evaluating infrastructure provisioning approaches. The participants were asked to carry out the experimental task as part of the laboratory exercises of the course.
Participants for this study will be recruited using a convenience sampling technique, a non-probability sampling method that selects individuals based on their availability and willingness to participate~\cite{Baltes:2022}. The study will target undergraduate and Master's students in computer science at the 
\textit{Universitat Politècnica de València}. 
%\textit{eliminated for double blind review}. 
We will will recruit participants during class time, providing a brief overview of the study and inviting interested students to participate. To ensure the voluntary nature of participation, we will emphasise that all participants have the opportunity to decline without any negative consequences. Once the participants have been selected, we will give them the informed consent document. The experiment will be conducted online and will follow the university IRB protocol. In particular, the consent form will state that no personal data is collected, or if it is collected, that it will not be published, and will be destroyed. 

\subsection{Design}

\begin{table}[]
\caption{The treatment sequence with three techniques (NA, HCI and HCI\&HF) and three application domains (Trips, Sports and Courses)}
\label{tab:crossover}
\begin{tabular}{
m{0.12\linewidth}
m{0.01\linewidth}
m{0.03\linewidth}
m{0.09\linewidth}
m{0.01\linewidth}
m{0.03\linewidth}
m{0.09\linewidth}
m{0.01\linewidth}
m{0.03\linewidth}
m{0.09\linewidth}
}
\hline
Domain    & \multicolumn{3}{c}{Trips}    & \multicolumn{3}{c}{Sports}   & \multicolumn{3}{c}{Courses}  \\
Period    & \multicolumn{3}{c}{Period 1} & \multicolumn{3}{c}{Period 2} & \multicolumn{3}{c}{Period 3} \\
Technique & NA     & HCI    & HCI\&HF    & NA     & HCI     & HCI\&HF   & NA     & HCI    & HCI\&HF    \\ \hline
Group 1   & \multicolumn{1}{c}{x}      & \multicolumn{1}{c}{-}      & \multicolumn{1}{c}{-}           & \multicolumn{1}{c}{-}       & \multicolumn{1}{c}{x}       & \multicolumn{1}{c}{-}      & \multicolumn{1}{c}{-}      & \multicolumn{1}{c}{-}       & \multicolumn{1}{c}{x}          \\
Group 2   & \multicolumn{1}{c}{x}      & \multicolumn{1}{c}{-}       & \multicolumn{1}{c}{-}           & \multicolumn{1}{c}{-}        & \multicolumn{1}{c}{-}        & \multicolumn{1}{c}{x}         & \multicolumn{1}{c}{-}       & \multicolumn{1}{c}{x}      & \multicolumn{1}{c}{-}            \\
Group 3   & \multicolumn{1}{c}{-}       & \multicolumn{1}{c}{x}      & \multicolumn{1}{c}{-}           & \multicolumn{1}{c}{x}      & \multicolumn{1}{c}{-}         & \multicolumn{1}{c}{-}           & \multicolumn{1}{c}{-}        & \multicolumn{1}{c}{-}        & \multicolumn{1}{c}{x}          \\
Group 4   & \multicolumn{1}{c}{-}        & \multicolumn{1}{c}{x}      & \multicolumn{1}{c}{-}            & \multicolumn{1}{c}{-}        & \multicolumn{1}{c}{-}         & \multicolumn{1}{c}{x}         & \multicolumn{1}{c}{x}      & \multicolumn{1}{c}{-}        & \multicolumn{1}{c}{-}           \\
Group 5   & \multicolumn{1}{c}{-}        & \multicolumn{1}{c}{-}       & \multicolumn{1}{c}{x}          & \multicolumn{1}{c}{x}      & \multicolumn{1}{c}{-}         & \multicolumn{1}{c}{-}           & \multicolumn{1}{c}{-}        & \multicolumn{1}{c}{x}      & \multicolumn{1}{c}{-}            \\
Group 6   & \multicolumn{1}{c}{-}        & \multicolumn{1}{c}{-}       & \multicolumn{1}{c}{x}          & \multicolumn{1}{c}{-}        & \multicolumn{1}{c}{x}       & \multicolumn{1}{c}{-}           & \multicolumn{1}{c}{x}      & \multicolumn{1}{c}{-}        & \multicolumn{1}{c}{-}            \\ \hline
\end{tabular}
\end{table}

    %The experiment will be designed as an AB/BA crossover design, which has one factor and two treatments. 
    The experiment is designed as a balanced within-subject three-treatment factorial crossover design.
    We will carry out a power analysis to allow us to determine the appropriate sample size to detect an effect of medium size.
    We chose this design because it addresses the issue of small sample sizes and increases the sensitivity of experiments.
    %where the Application Domain (adaptive system) is applied in two periods and (see Table 1).
    In a balanced crossover design the measures are taken several times from a participant (i.e., a participant is assigned to a sequence of treatments). We follow the guidelines proposed by Vegas et al. \cite{vegas:2015} to define the crossover design. In particular, we will analyse not only the effects of the treatments (i.e., NA, HCI and HCI\&HF) on the dependent variables, but also the effects of critical crossover variables (i.e., period, sequence and carryover). %A period is defined by the application of one treatment by one participant to one Application Domain (adaptive system). 
    In this scenario, it is necessary to differentiate between the concepts of period and session. A period is defined by the application of one treatment by one participant to one application domain (adaptive system), whereas a session is a portion of time taken by a subject to complete (one or more) experimental tasks~\cite{vegas:2015}. We consequently will have three periods, since each participant had to perform two treatments and for reasons of the students’ class timetable. We, therefore, will carry out one period in a session.
    Carryover is the persistence of the effect of one treatment when another treatment is applied later. The objective is to find out whether these additional factors are influencing the dependent variables. Table \ref{tab:crossover} shows that we have a special type of design called a factorial crossover design, which has the same number of periods as treatments.

    %, in which all the participants apply each treatment under study once and once only~\cite{vegas:2015}.  
    %We chose the AB/BA crossover design because it addresses the issue of small sample sizes and increases the sensitivity of experiments. 

    %In this context, the reward model employed to adapt the user interface (HCI or HCI\&HF) is the factor, and the two treatments are the \textit{Courses management system 
    %catalogue} and the \textit{Sporting goods store
    %Sports catalogue}. 

    %We, therefore, will use period, sequence and carryover as fixed factors.  
    %Esto se debería decir en la sección de análisis de datos.

    Each period will be assigned to a domain and each group will follow a different sequence of adaptation techniques. Producing 6 different sequences (see Table \ref{tab:crossover}).
    In the first period, every group will interact with a Trip Planner system. Groups 1 and 2 will use the system without adaptation capabilities (NA), while Groups 3 and 4 will interact with AUI that employs a RL agent with predictive HCI models (HCI) and Groups 5 and 6 will interact with AUI that employs a RL agent with predictive HCI models and Human Feedback (HCI\&HF).
    In the second period, everyone will interact with an e-commerce system (Sporting Goods Store). Groups 3 and 5 will use the system without adaptation capabilities (NA), while Groups 1 and 6 will interact with AUI-HCI and, Groups 2 and 4 will interact with AUI-HCI\&HF.
    In the third period, everyone will interact with an e-learning system (Courses Management system). Groups 4 and 6 will use the system without adaptation capabilities (NA), while Groups 2 and 5 will interact with AUI-HCI and, Groups 1 and 3 will interact with AUI-HCI\&HF.
    We carefully selected these sequences to ensure that any potential order effects are minimised. In this scenario, we do not believe that there is the possibility that any of the sequences would have improved the experimental results.

    % The experiment will also include a Baseline period, which will occur before the first period. The Baseline period will be used to establish the participants' initial level of performance without any intervention. During this period, all participants will use the non-adaptive UI (i.e., the control condition) with a Trip Planner system. The purpose of the Baseline period is to establish a performance baseline against which the performance in the subsequent periods can be compared.    
    % After the Baseline period, since we had two periods and two treatments, there were two resulting sequences, that is, Baseline-Sports-Courses and Baseline-Courses-Sports. 
    % In the first period, Group 1 will interact with AUI that employs a RL agent with predictive HCI models (technique HCI) in an e-commerce system (Sporting Goods Store), while Group 2 will use the same RL technique with
    % an e-learning system (Courses Management system).
    %a Courses Management system. 
    % In the second period, Group 1 will interact with AUI that employs a RL agent with predictive HCI models and HF (technique HCI\&HF) in the e-learning system, while Group 2 will use the same RL technique with the e-commerce system. We carefully selected these sequences to ensure that any potential order effects are minimised. In this scenario, we do not believe that there is the possibility that any of the sequences would have improved the experimental results.

    We choose the trip, sporting, and courses systems from the e-commerce and e-learning domains as application domains for this experiment because, these domains are representative of different types of software products and user requirement needs. Moreover, these domains are popular and widely used, making them relevant for a wide range of users. Finally, the selected systems have similar complexity.
    %, making them suitable for testing the effectiveness of RL-based adaptive UIs.

%In the first period, participants will use the AUI-baseline with a travels catalogue. This baseline will serve as a reference allowing us to capture the user engagement where no adaptations are provided by the UI. 

\subsection{Analysis plan}

The results of the experiment will be collected using \textit{i)} the user engagement computed with the predictive HCI models exclusively and \textit{ii)} predictive HCI models with HF.
Additionally, the participants will carried out a post-experimental task that we will allow us to measure their user satisfaction by using the QUIS questionnaire.

%with the adaptive and non-adaptive   Hand;\textit{ii)} the SUS questionnaire responses.
We will use descriptive statistics, violin plots, and statistical tests to analyse the data collected from the experiment. As is usual, we accepted a probability of 5\% of committing a Type-I Error in all the statistical tests. 

% \begin{itemize}
%     \item 
    We will first carry out a descriptive study of the measures for the dependent variables. Then,
    % \item 
    as a crossover design will be used in the study, it is necessary to analyse the experiment factors including periods, sequences, and carryover. To test the formulated hypotheses, the Linear Mixed Model (LMM) will be employed. Following the guidelines proposed by Vegas et al. \cite{vegas:2015}, the LMM includes the following fixed factors: Reward model technique (treatment), period (confounded with the application domain), and sequence (confounded with carryover and period-technique interaction), and subject as a random factor nested within the sequence. The LMM will be used to assess whether these factors have influenced the results. All the assumptions of LMMs will be tested and reported. In order to apply the LMM, the residuals had to meet the condition of normality~\cite{vegas:2015}. To ensure that the model was valid, we therefore will use the Shapiro-Wilk test if the sample size ends up being smaller than 50 and Kolmogorov–Smirnov if the sample size is higher than 50 to confirm the normality of the residuals.
    We will then apply the LMM to each dependent variable in order to assess whether the period (confounded with the application domain), sequences (confounded with period-technique and carryover) or technique (treatment) had statistical significance.
    %
    
    % \item 
    We will report the results of the QUIS scores for the experiment. We will also assess the questionnaire quality in terms of its reliability, validity and sensitivity.
    Then, 
    % \item 
    %since the statistical significance is not sufficient to explain the difference caused by the treatments, 
    we will measure the effect size to assess the magnitude of differences caused by the treatments. The effect size of the treatments should be measured only if the period, sequence or any blocking variables have no bearing, and there is no carryover~\cite{carver:2010}. To this end, we will use Cohen's d. 
    % We will interpret the results using the guidelines: \textit{$d<0.2$} (very small),\textit{$0.2<d<0.5$} (small) \textit{$0.5<d<0.8$} (medium) and \textit{$0.8<d$} (large).
    
% \end{itemize}

\subsection{Threats to validity}

In this section, we 
%follow the recommendations of Wohlin et al.~\cite{Wohlin:2012} 
discuss the issues that will may threat the validity of our experiment and how we will mitigate them ~\cite{Wohlin:2012}.
One potential threat to internal validity is the Hawthorne effect~\cite{adair1984hawthorne}, where participants may alter their behaviour due to being observed. To mitigate this threat, participants will be instructed to behave as they would in a natural setting. Another potential threat is the maturation effect, where participants' responses may change over time due to natural changes such as fatigue or boredom. To mitigate this, we will have a 
%short 
long break in the middle of each session. % to allow participants to rest.
%
% R1. Comment 6. Interaction between groups
%
Also, to mitigate potential threats related to participant's interaction, participants will be recruited from different academic degrees (i.e., Bachelor's and Master's in Computer Science) which happen in different semesters. Participants from these groups are not expected to interact with each other during the study periods/sessions.
% This deliberate design choice aims to minimize the potential for cross-group interaction, thereby maintaining the internal validity of the study by controlling for external influences that could confound the results.

To address the potential threat to conclusion validity, a sufficient number of participants will be recruited to ensure adequate statistical power and the proper statistical tests will be performed for the variables and their assumptions~\cite{Maxwell2002Applied}.

The potential threats to the construct validity, are the validity and reliability of the HCI and HCI\&HF models and the questionnaires used to measure user satisfaction and engagement. To mitigate these threats, the models will be based on existing research~\cite{LEHMANN:2012, barbaro:2020, carlton:2021} and validated by domain experts. 
To validate the measures of user engagement obtained with the predictive HCI models, during the interaction, we will gather the participants' engagement through User Engagement Scale (UES)~\cite{Obrien:2018} and the participants' satisfaction through QUIS questionnaire~\cite{norman:1998}, which are empirically validated questionnaires, as post-experimental tasks. 
%Moreover, empirically validated questionnaires (i.e., SUS and UES) will be used to ensure the reliability of capturing user satisfaction and engagement.

%Additionally, the tasks used to measure the effectiveness of the adaptive UIs will be designed to reflect realistic browsing and purchasing behaviors.

% One potential threat to construct validity is the validity of the HCI and HCI\&HF models used in the experiment. To increase the construct validity, the models will be designed based on existing research~\cite{LEHMANN:2012, barbaro:2020} and validated by domain experts.
% Another potential threat to construct validity is the reliability of the questionnaires to capture the user satisfaction and user engagement. To mitigate this
% Finally, the validity of the tasks used to measure the effectiveness of the adaptive UIs could affect the construct validity. To mitigate this effect, the tasks will be designed to reflect realistic browsing and purchasing behaviours. 

To increase external validity, we will address the potential threat of participants' behaviour not generalising to real-world situations by using a realistic scenario for each domain to make the tasks more meaningful. Additionally, we will design the tasks used to measure the effectiveness of the adaptive UIs to reflect realistic browsing and purchasing behaviours.
Nevertheless, the representativeness of the results may be affected by the profile of the participants. Replications with industry participants are necessary.
% One potential threat to external validity is the generalizability of the findings. The participants' behaviour in the study may not generalise to real-world situations. To mitigate this threat, we will use a realistic scenario for each domain to make the tasks more meaningful to the participants. 
% The tasks used to measure the effectiveness of the adaptive UIs will be designed to reflect realistic browsing and purchasing behaviours. 
% This will help to ensure that participants are engaged in the task and that the results are applicable to real-world situations. 

\section{Execution Plan}

The experiment will be conducted over the course of two weeks, with each participant attending three sessions.
% Fig. \ref{fig:ExecutionPlan} b) illustrates the experimental design for each group.
Before the first session, all participants will complete a brief demographic survey and will be assigned to one of the six groups.
% and a baseline task using the AUI-baseline on a Trip Planner system.
In the first session, they will interact with a Trip Planner system using the three approaches (NA, HCI, HCI\&HF).
Groups 1 and 2 will use the system without adaptation capabilities (NA). Groups 3 and 4 will interact with AUI-HCI and, Groups 5 and 6 will interact with AUI-HCI\&HF.
In the second session, participants will switch to another approach in an e-commerce system (Sporting Goods Store). Specifically, Groups 3 and 5 will use the NA, while Groups 1 and 6 will use the HCI and, Groups 2 and 4 will use the HCI\&HF. 
Finally, in the third session, participants will switch to the last approach that they haven't used in an an e-learning system (Courses Management system). Groups 4 and 6 will use NA, while Groups 2 and 5 will use HCI and, Groups 1 and 3 will use HCI\&HF.

% Group 1 will use the HCI with the Sporting Goods Store, while Group 2 will use the HCI\&HF with the Courses Management system. Participants in each group will complete a set of predefined tasks related to browsing and selecting items in catalogues. 
% In the second session, participants will switch to the opposite adaptive UI and domain combination. Specifically, Group 1 will use the HCI\&HF with the Courses Management system, while Group 2 will use the HCI\&HF with the Sporting Goods store. Again, participants in each group will complete the same set of tasks. 
%as in the first session.
Each session will consist of the following steps:

\begin{enumerate}
    \item Introduction and task explanation.
    %: The experimenter will introduce the session and explain the task to the participant.
    \item Task execution: Participants will perform the tasks using the assigned AUI. The tasks will be defined once the 
    %alternative 
    adaptations for the selected sytems are defined.

    % \begin{itemize}
    %     \item Trip Planner system (control): The experimenter will ask the participants to imagine that they are planning a trip for 5 days with a limited budget. They will have to explore the system looking for the best option.
        
    %     \item For the Sporting Goods store: The experimenter will ask the participants to imagine that they are preparing for a marathon. The participants will have a limited budget but they will have to add as many items as possible. To avoid confusion or uncertainty, the experimenter will give a brief clarification on what type of items are relevant for a marathon.
        
    %     \item For the Courses management system: The experimenter will ask the participants to imagine that they are interested in taking online courses to improve their skills in a specific area. 
    %     %programming. 
    %     The participants will have a limited budget but they will have to add as many courses as possible. To avoid confusion or uncertainty, the experimenter will give a brief clarification on what type of courses are relevant for that specific area.
    %     %programming
        
    %     %\item For the Books catalogue: The experimenter will ask the participants to imagine that they are planning a trip to a foreign country and they want to learn as much as possible about the local culture, history, and language. The participants will have a limited budget but they will have to select and add as many books as possible to their chart to prepare for the trip.
    % \end{itemize}
    
    \item Post-task questionnaires: The participant will be asked to fill the QUIS and the UES to evaluate their satisfaction and engagement with the UI, respectively. The questionnaires will be administered through an Excel file and participants will have to complete after each session.
\end{enumerate}

All sessions will be conducted in a controlled lab environment with identical equipment and software configurations. Participants will be seated at a desk with a computer screen, mouse, and keyboard. The order of the sessions will be counterbalanced to control for any potential order effects. 
Each session will last, at most, 120 minutes. However, participants will not be required to complete the tasks within this time frame, and there will be no time pressure on them to do so. 
%During the sessions, participants' interactions with the UI will be recorded using screen capture software.

\section*{Acknowledgment}
This work is supported by the AKILA project (CIAICO/2021/303)
funded by the Generalitat Valenciana (GVA). Daniel Gaspar-Figueiredo is funded by the GVA under the grant ACIF/2021/172, which is cofunded by the European Union through the ESF.

\bibliographystyle{IEEEtran}
\bibliography{main}

\end{document}